\documentclass[preprint,12pt]{elsarticle}

\usepackage{amssymb}
\usepackage{lipsum}
\usepackage{graphicx}
\usepackage{siunitx}

\journal{qqq $\&$ Computing}

\begin{document}

\begin{frontmatter}



\title{A Real-time Non-contact Localization Method for Faulty Electric Energy Storage Components using Highly Sensitive Magnetometers}


\author[one]{Tonghui Peng}
\author[one]{Wei Gao*}
\ead{weigao@ustc.edu.cn}
\author[one]{Ya Wu}
\author[three]{Yulong Ma}
\author[one]{Shiwu Zhang}
\author[two]{Yinan Hu*}
\ead{ynhu@ibp.ac.cn}
\affiliation[one]{organization={CAS Key Laboratory of Mechanical Behavior and Design of Materials, 
Department of Precision Machinery and Precision Instrumentation,
School of Engineering Science, University of Science and Technology of China},
            city={Hefei, Anhui},
            postcode={230026}, 
            country={China}}
\affiliation[two]{organization={State Key Laboratory of Brain and Cognitive Science, Institute of Biophysics, Chinese Academy of Sciences},
            city={Beijing},
            postcode={100101}, 
            country={China}}
\affiliation[three]{organization={Chinainstru \& Quantumtech (Hefei) Co., Ltd, Hefei, China},
            city={Hefei, Anhui},
            postcode={230031}, 
            country={China}}
            
\begin{abstract}



With the wide application of electric energy storage component arrays, such as battery arrays, capacitor arrays, inductor arrays, their potential safety risks have gradually drawn the public attention. However, existing technologies cannot meet the needs of non-contact and real-time diagnosis for faulty components inside these massive arrays. To solve this problem, this paper proposes a new method based on the beamforming spatial filtering algorithm to precisely locate the faulty components within the arrays in real-time. The method uses highly sensitive magnetometers to collect the magnetic signals from energy storage component arrays, without damaging or even contacting any component. The experimental results demonstrate the potential of the proposed method in securing energy storage component arrays. Within an imaging area of 80 mm $\times$ 80 mm, the one faulty component out of nine total components can be localized with an accuracy of 0.72 mm for capacitor arrays and 1.60 mm for battery arrays.

\end{abstract}




\end{frontmatter}


\section{Introduction}
\label{introduction}


Electric energy storage components can be used for storing electric energy and releasing it for power when needed, such as batteries~\cite{van2014better}, capacitors~\cite{nishino1996capacitors,nomoto2001advanced}, and inductors~\cite{mclyman2004transformer}. These components play pivotal roles in contemporary society, contributing to various applications and advancements. However, the energy and power density of a single energy storage component is often limited. To meet the needs for large-scale energy storage, energy storage components usually exist in the form of arrays. For example, there are more than seven thousands of cells (18650 battery) inside the battery pack of the Tesla Model S~\cite{saw2016integration}. However, fault within any single cell will accelerate the aging of the whole battery pack~\cite{feng2019propagation} and eventually lead to various car accidents~\cite{sun2020review,liao2019survey}. Therefore, early identification and localization of malfunctioning components within the array are crucial tasks.

Current diagnostic methods for faulty energy storage components can be classified into four categories: model-based methods~\cite{chen2016model}, signal-processing-based methods~\cite{kang2019multi}, data-driven methods~\cite{xue2021fault}, and knowledge-based methods~\cite{li2014capacity}. Regardless of the method chosen for fault diagnosis, it is necessary to collect the voltage, current, or temperature signals of each individual cell within the battery pack, which remains challenging to be accomplished in an economic and timely manner.

In recent years, it has become possible to measure the magnetic field around the battery pack for non-destructive and non-contact diagnosis~\cite{hu2020sensitive,hu2020rapid}. Compared with traditional voltage and current diagnosis, magnetic diagnosis does not need to contact the battery pack and will not affect the electrochemical process inside. Compared with traditional temperature diagnosis, magnetic diagnosis is more accurate and responsive. Nevertheless, implementing magnetic field detection still remains challenging. Recent advancements in magnetic sensing technology have led to the development of highly sensitive and reliable magnetic field sensors, such as optically pumped magnetometer (OPM)~\cite{allred2002high}. Therefore, this work has adopted an OPM as the magnetic field detection device, taking advantage of its superior sensitivity and accuracy in detecting magnetic fields. The OPM works on the principle of optical pumping to realize highly sensitive magnetic measurement with high-pressure vapor. In the process, the vapor is optically pumped into a magnetically sensitive state, which would be disrupted by the presence of magnetic fields, resulting in detectable voltage changes on a photodiode. This enables OPM to measure extremely weak magnetic fields with high sensitivity~\cite{tierney2019optically}. OPM has been widely used for imaging localization of biomagnetic fields such as brain and heart~\cite{boto2017new,alem2015fetal}.

With the acquisition of high-precision magnetic field data using OPM, corresponding spatial filtering algorithms are required to extract the signals generated by fault sources from the magnetic field data. The beamforming algorithm, serving as a practical spatial filtering technique, has been widely used in array signal processing. By synthesizing the signals received from multiple measurements, it can enhance the signals in specific directions and weaken the interference signals in other directions, thereby improving the quality and reliability of the desired signals. This algorithm is often used in noise positioning~\cite{chiariotti2019acoustic}, language enhancement~\cite{gannot2001signal}, 5G communication~\cite{roh2014millimeter}, biomagnetic positioning~\cite{hillebrand2005beamformer}, etc.

This work makes a significant contribution by overcoming the limitations of traditional diagnostic methods (contact required, less accuracy, slower response, etc) and realizing rapid and precise localization of the faulty component in an energy storage component array. Only a single OPM is used to conduct non-destructive and non-contact fault detection inside the array to avoid the interference when multiple OPMs are deployed. The beamforming algorithm is then  applied to extract the precise location of the faulty component in real-time out of the raw outputs from the OPM. To the best of our knowledge, it is the first time that the method combining OPM and the beamforming algorithm is applied to the diagnosis of energy storage component arrays.

\begin{figure}
    \centering
    \includegraphics[width=14cm]{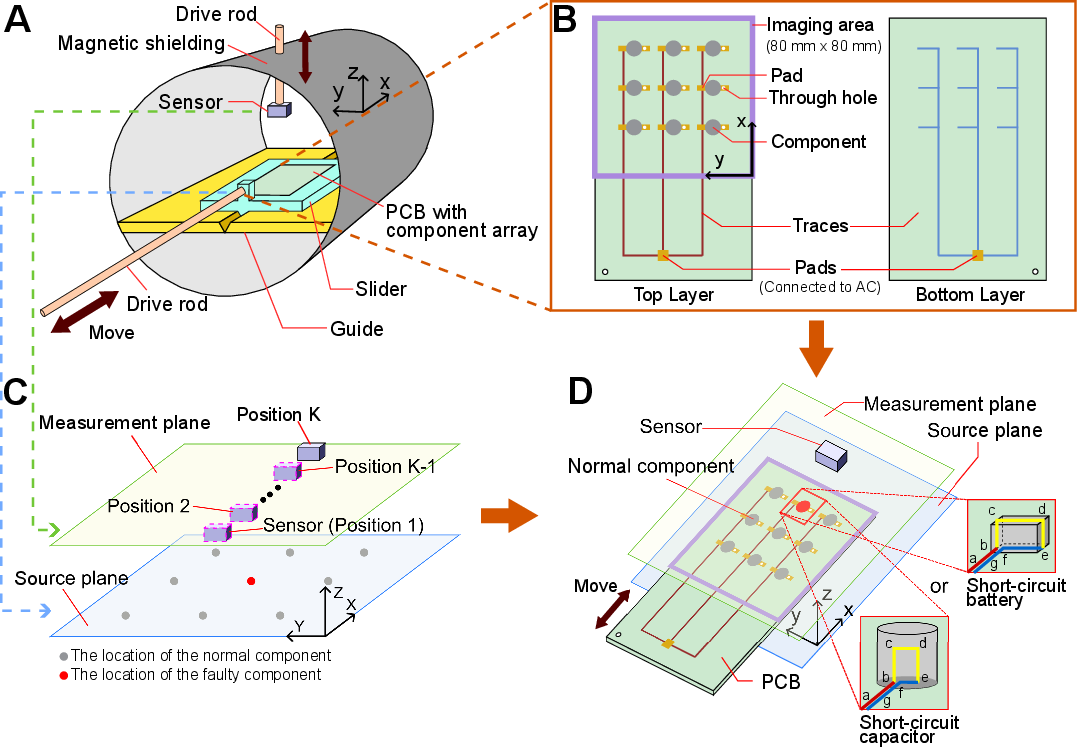}
    \caption{Overview. (A) The experimental arrangement. The energy storage component array to be diagnosed exist on a Printed Circuit Board (PCB). A stepper motor drives the transmission rod to move the slider back and forth along the x-axis direction while the PCB is fixed to the slider. To avoid external magnetic interference and satisfy the normal working conditions of the OPM, the entire experiment is run inside a magnetic shielding cylinder. Note that only the innermost shield is shown, whereas the other six concentric cylinders are hidden for brevity. (B) The top and bottom views of the PCB. (Left) PCB top layer structure. Nine energy storage components are connected in parallel on the PCB. (Right) PCB bottom structure. The bottom structure is almost completely symmetrical to the top layer. The pads on both top and bottom layers are connected to external AC source to charge and discharge all the components at the same time. (C) The source and measurement planes. The red and gray dots on the source plane represent the components, whereas the block on the measurement plane represents the sensor, including K measuring locations.(D) A combined demonstration of the PCB, the source plane and the measurement plane. During the experiment, the PCB carrying the energy storage component array was moved along the x-axis on the source plane, and the generated magnetic field was recorded by a sensor on the measurement plane. (Inset) Illustration of a component (capacitor or battery) with a short-circuit fault. b-c-d-e (yellow) is the short-circuit inside the component, a-b (red) is the circuit on the top layer of the PCB, and e-f-g (blue) is the circuit on the bottom layer of the PCB.}
    \label{fig:1}
\end{figure}

During the detection process in this work, the OPM is placed above the energy storage component array, or the source plane, and measures the mixed signals from both normal and faulty components below. Traditional detection methods involves scanning the entire source plane with sensors to identify points with abnormal signals, which not only takes time but also lacks precision. In this paper, the proposed method requires only a few evenly spaced measurements of the magnetic fields above the source plane, which is realized by moving the source plane along the x-axis via a guiding rail, as shown in Figure~\ref{fig:1}. It can be seen that, to cancel external magnetic field interference, the OPM is fixed on the measurement plane and operated within a magnetic shield. As the source plane is moving, the relative position changes between the array and the OPM, resulting in raw magnetic field data from specific locations. To implement the beamforming algorithm, the entire source plane is divided into evenly distributed grids, and then the beamforming algorithm converts the measured raw data back into the original signals from each grid point. The grid points with abnormal signals can then be identified and thus the location of the faulty component.

\section{Simulation}

To simulate the working environment, the energy storage components are connected in parallel on a printed circuit board (PCB) and an external signal source is used to charge and discharge the component array, as shown in Figure~\ref{fig:1}(B). The energy storage component array consists of eight normal components and one faulty component, where the faulty component is short-circuit like in the most commonly occurring faulty state, as shown in Figures~\ref{fig:1}(C) and~\ref{fig:1}(D). During simulation, the magnetic fields of the energy storage components are simulated on the source plane using the magnetic source model. The difference between normal and faulty storage components lies in the phase difference of the current. Taking two capacitors as an example, the normal capacitor exhibits capacitance characteristics with the current lagging the voltage by 90 degrees, while the faulty capacitor mainly exhibits resistance characteristics, with the current in phase with the voltage. On the measurement plane, a sensor collects raw magnetic field data at four different positions. The magnetic field measured at the sensor is the superposition of magnetic fields generated by the currents with two different phases. The focus of this simulation is to establish the corresponding model and simulate the acquisition of magnetic field data by a magnetometer. It should be noted that the data contain the total magnetic field generated by all components and require the beamforming algorithm to filter and extract the actual location of the faulty component.

The following subsections present more details about the simulation setup and the corresponding results.

\subsection{Simulation Model}

Previous work has demonstrated that the magnetic dipole model can be utilized to describe the magnetic field distribution of batteries in a non-operational state~\cite{hu2020sensitive}. However, in this work, it is necessary to measure the magnetic field signals of energy storage components during the charging and discharging process, thus requiring the development of a new model.

When a short-circuit fault occurs in the energy storage component (as shown in Figure~\ref{fig:1}(D)), the current inside the component will flow from one pin to the short-circuit point and then to the other pin (segments b-c-d-e in the illustration), and finally flow away on the back of the board (segment e-f in the illustration). Therefore, Biot-Savart's law can be used to establish the relationship between the current through the component and the magnetic field around the component as 
\begin{equation} \label{eq1}
 \mathbf{b}_{k,\theta} = \frac{\mu_{0}I_{\theta}}{4\pi} \int_L \frac{\text{d}\mathbf{l} \times \mathbf{r}_{k,\theta}}{|\mathbf{r}_{k,\theta}|^{3}}= \mathbf{h_{k,\theta}}\cdot I_{\theta},
\end{equation}\par
\noindent where $\mu_{0}$ is the vacuum permeability, $\theta = (x, y, z)$ represents a position on the source plane, $k = 1,2,...,K$ is the position index on the measurement plane, $I_{\theta}$ stands for the source current at position $\theta$, $L$ is the integral path, $\text{d}\mathbf{l}$ is a differential element of the integral path, $\mathbf{r}_{k,\theta}$ is the vector from position $\theta$ pointing to the $k$-th measurement position, $\mathbf{h_{k,\theta}}$ describes relationship between the source current $I_{\theta}$ on the source plane and the $k$-th measured magnetic field $\mathbf{b}_{k,\theta}$ on the measurement plane. 

It should be emphasized that the magnetic dipole model (as shown in Equation~(\ref{eq1})) can be used to describe the magnetic field generated by both normal and faulty energy storage components, because the volume current and displacement current~\cite{hyodo2022maxwell} inside the normal energy storage components can also be equivalent to the loop shown in Figure~\ref{fig:1}(D). The only difference between the magnetic fields generated by normal and faulty component is the phase of the alternating magnetic field, which is critical for locating the faulty component using the beamforming algorithm in the subsequent steps.
Considering $M$ energy storage components located on a source plane, with each component's magnetic field described by Equation~(\ref{eq1}), and $K$ measurement positions on a measurement plane, the problem can be formulated as a multi-input multi-output system, 
\begin{equation} \label{eq2}
      \mathbf{B}=\mathbf{H}\mathbf{Q} + \mathbf{V},
\end{equation}\par
\noindent where the matrix $\mathbf{B}$ combines the magnetic fields measured at all positions on the measurement plane at $N$ different time points, the matrix $\mathbf{Q}$ represents the current flowing through the energy storage components on the source plane, the matrix $\mathbf{H}$ is the so-called lead field matrix which maps the source current distribution to the magnetic field distribution, the matrix $\mathbf{V}$ represents the measurement noise or any other external interference. Note that, the elements inside the matrix $\mathbf{B}$ can be expressed as $\mathbf{B}_{kn}=\sum\nolimits_{m=1}^{M}\mathbf{b}_{k,\theta_{m}} (\text{where}\ t=t_n)$, which represents the magnetic field vector collected by the sensor at the $k$-th location and the $n$-th time instance. This relationship connects Equation~\ref{eq2} with Equation~\ref{eq1}. The ultimate objective is to solve the inverse problem of obtaining the current signal, and hence locate the faulty component on the source plane, using the measured magnetic field signals and the actual sensor positions as inputs to Equation~(\ref{eq2}). In the next subsection, the beamforming algorithm will be utilized to address this inverse problem.

\subsection{Beamforming}
Beamforming algorithm is a widely used technique in signal processing for spatial filtering and source localization. It filters the measured signals to extract the signals from the desired source location by setting spatial filters in the source space. The Linearly Constrained Minimum Variance (LCMV)~\cite{baillet2001electromagnetic} method can be employed to design spatial filters. The core idea of LCMV method is to optimize the weights of the filter such that the signal at non-desired locations is suppressed to zero.
Consequently, the source activity strength (estimated variance of current source) at location $\theta$ on the imaging plane can be obtained as~\cite{hillebrand2005beamformer}
\begin{equation} \label{eq3}
    \sigma _{\theta }^{2}=\left ( \mathbf{h}_{\theta }^{T}\mathbf{C_{B}}{\mathbf{h}_{\theta }}\right )^{-1},
\end{equation}\par
\noindent where $\mathbf{h}_{\theta }$ is a column of the leaf field matrix $\mathbf{H}$ corresponding to the position $\theta$, $\mathbf{h}_{\theta} = \left [ \mathbf{h}_{1,\theta}^{T},\mathbf{h}_{2,\theta}^{T},...,\mathbf{h}_{K,\theta}^{T}\right ]^{T}$, which can be calculated by Equation~(\ref{eq1}), $\mathbf{C_B}$ is the covariance matrix of the magnetic field signal $\mathbf{B}$.

The source plane is divided into many grids (each grid size is $0.1 mm\times0.1 mm$ ). Using Equation~(\ref{eq3}), the source strength at each grid can be reconstructed, and ultimately an imaging map of the magnetic field on the source plane can be obtained. The beamforming algorithm requires zero correlation among source signals, but in the source plane, the $M-1$ normal storage components have high correlation, making imaging impossible. As a result, only the faulty component is successfully imaged, thus achieving localization of the faulty component.

Equation~(\ref{eq3}) treats the noise signal as if it were a source imaging on the imaging plane, which causes the unevenness of the imaging plane. However, under the assumption of known prior information on sensor noise, the normalized source estimate intensity output can be written as~\cite{robinson1999functional}
\begin{equation} \label{eq4}
z _{\theta }^{2}=\left ( \mathbf{h}_{\theta }^{T}\mathbf{C_{B}}{\mathbf{h}_{\theta }}\right )^{-1}/\left ( \mathbf{h}_{\theta }^{T}\mathbf{C_{V}}{\mathbf{h}_{\theta }}\right )^{-1},
\end{equation}\par
\noindent where $\mathbf{C_{V}}$ is an estimate of the noise-only covariance, $z _{\theta }^{2}$ is the normalized source activity intensity called the source activity index. 

In the next section, the results of using this method to locate the faulty storage component in simulation will be presented. The differences in imaging results using two indicators, i.e., the source activity intensity in Equation~(\ref{eq3}) and the source activity index in Equation~(\ref{eq4}), will be presented.

The source activity intensity is the estimated variance of the original signal, which reflects the power of a source at a specific position on the imaging plane. A value close to 0 indicates the absence of a target source (faulty storage component) at that position.

The source activity index is the normalized version of the source activity intensity. It represents the ratio of the power of the original signal to the power of the noise signal at a particular position. It is a dimensionless quantity, and a higher value indicates the presence of a target source (faulty storage component) at that position.

\subsection{Simulation Result}
\subsubsection{Locating a faulty source}

\begin{figure}
    \centering
    \includegraphics[width=14cm]{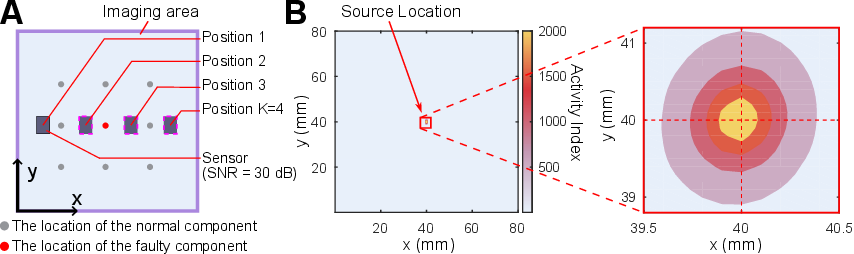}
    \caption{Top view of the simulation setup and the imaging results of fault location in the simulation environment. During the simulation process, the sensor's signal-to-noise ratio (SNR) is maintained at 30 dB. (A) The red dot represents the position of the faulty component to be located. Magnetic field signals are collected by one sensor at four different positions. (B) Imaging result of fault location. The red arrow indicate the location of the faulty component. The colors in the plot represent the activity index of the imaged source, which is the ratio of the estimated intensity of the source to the estimated intensity of the noise. (Inset) Enlarged view of the imaging of the faulty component.}
    \label{fig:2}
\end{figure}

The purpose of this simulation is to demonstrate that our method is capable of accurately localizing the faulty component in an energy storage component array. The top view of the simulation setup is shown in Figure ~\ref{fig:2}(A). Nine components are placed on the source plane, where the red one represents a faulty component (located at coordinates (40.0 mm, 40.0 mm)) and the gray ones are normal components. The charging and discharging frequency of the energy storage components is 2 Hz. Gaussian white noise with a certain magnitude is actively added during the simulation process to realize a signal-to-noise ratio (SNR) of 30 dB for the measured magnetic field.

Figure~\ref{fig:2}(B) shows the imaging results of the faulty component localization. It can be seen that the only one source is located, with a center coordinates of (40.0 mm, 40.0 mm). This location precisely corresponds to the position of the faulty component in Figure~\ref{fig:2}(A). The localization accuracy is within 0.1 mm. The zoomed-in image, which is shown at the right panel in Figure~\ref{fig:2}(B), reveals detailed information about the localization results. It can be observed that, under a signal-to-noise ratio of 30 dB, the faulty component localization is not only of aforementioned high accuracy, but also of high resolution, with a value of less than 0.5 mm.

When implementing beamforming algorithms for signal processing, it is crucial to ensure that the time-during of the sources are not highly correlated. In situations where the time-during are highly correlated, the beamforming algorithm may treat multiple correlated signals as a single signal, which can hinder effective signal separation and identification. This feature has been successfully taken advantage of in the imaging process, where the perfect correlation of the magnetic moments of the gray magnetic dipoles for normal components in Figure~\ref{fig:2}(A) prevents them from being properly imaged at the imaging plane, as shown in Figure~\ref{fig:2}(B). In reality, the majority of the components in arrays are normal ones that cannot be imaged using this technique because of their magnetic coherence. However, the faulty component can be imaged due to their phase difference. This enables the identification and localization of the faulty component out of the array, which also highlights the effectiveness and reliability of the beamforming algorithm for fault detection and diagnosis.

\subsubsection{Comparison of Two Imaging Indicators}

The imaging results in Figure~\ref{fig:2} were obtained under the condition of a signal-to-noise ratio of 30 dB. However, in working environment, the accuracy of fault localization may be affected by the background noise of the sensor. Therefore, it is necessary to investigate the performance of fault localization and resolution under such circumstances.

\begin{figure}
    \centering
    \includegraphics[width=14cm]{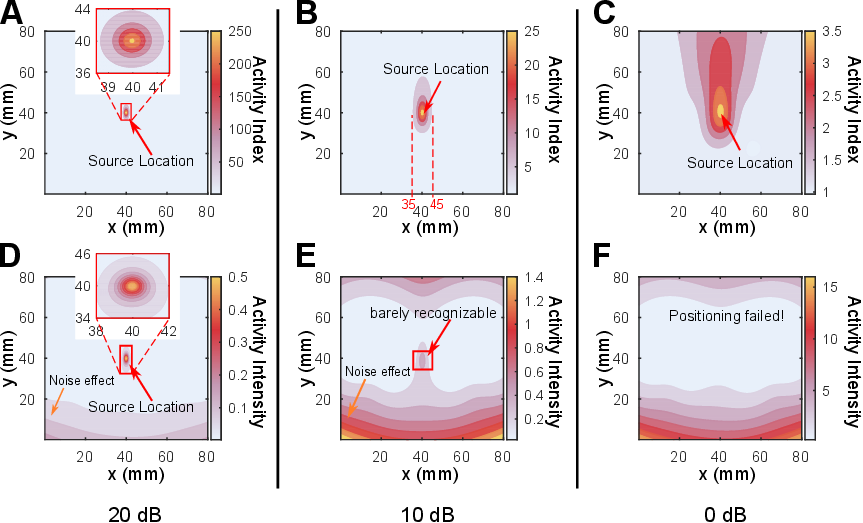}
    \caption{The imaging maps corresponding to the two imaging indicators under different signal-to-noise ratio (SNR) conditions in the simulation environment. The corresponding sensor signal-to-noise ratio is given at the bottom where the noise is manually added Gaussian white noise under simulation conditions. The source of the imaging is a faulty component in an energy storage array. The settings of the sources and sensor are shown in Figure~\ref{fig:1}(C) and Figure~\ref{fig:2}(A). (First row) Imaging maps using the indicator of source activity index. (Second row) Imaging maps using the indicator of source activity intensity. The difference between these two imaging indicators is that the source activity index takes into account the impact of sensor noise on imaging results.}
    \label{fig:3}
\end{figure}

Figure~\ref{fig:3} shows the differences in imaging maps under different signal-to-noise ratios and different imaging indicators. Figures~\ref{fig:3}(A),~\ref{fig:3}(B) and~\ref{fig:3}(C) show the performance of the imaging maps under different signal-to-noise ratios using the indicator of source activity index. These three sub figures all show that the locations of the fault source have been successfully located. However, as the signal-to-noise ratio decreases, the area of the location source expands, which indicates that the resolution of the faulty component's location in the array decreases. The fault localization results are represented by the coordinate values in the form of (x, y), which can then be compared with the actual coordinates of the fault source, (40.0 mm, 40.0 mm). It can be seen that the fault localization results are (40.0 mm, 40.0 mm) for the signal-to-noise ratio of 20 dB as in Figure~\ref{fig:3}(A), (40.1 mm, 40.1 mm) for 10 dB as in Figure~\ref{fig:3}(B), and (40.6 mm, 40.2 mm) for 0 dB as in Figure~\ref{fig:3}(C). 
These results demonstrate that the accuracy of fault localization is highly dependent on the signal-to-noise ratio, with lower signal-to-noise ratios corresponding to lower accuracy.

Figures~\ref{fig:3}(D),~\ref{fig:3}(E) and~\ref{fig:3}(F) show the performance of the imaging maps under different signal-to-noise ratios using the indicator of source activity intensity. The advantage of this imaging indicator is that the resulting image represents the signal strength of the source. However, the disadvantage is that the noise is also imaged on the imaging plane, meaning that non-localizing points on the imaging plane can also be reconstructed to have signal strength, as indicated by the orange arrows in Figures~\ref{fig:3}(D) and~\ref{fig:3}(E), which can affect the determination of the fault location. Additionally, when the signal-to-noise ratio is low as shown in Figure~\ref{fig:3}(F), fault imaging may fail. In conclusion, too low signal-to-noise ratio will seriously reduce the positioning accuracy and the imaging resolution. At the same time, using the indicator of source activity index is better in imaging performance, which is used for imaging in the subsequent experiments.

\section{Experiment on Fault Localization of Energy Storage Component Arrays}

The simulation results have demonstrated the effectiveness of the proposed method in locating the faulty component in an energy storage array. To validate the performance of the proposed method in practical scenarios, two experiments were designed to locate the faulty component in a capacitor array and a battery array, respectively. The experimental results are compared with the simulation results to evaluate the method's robustness and accuracy. 

\subsection{Experimental setup}

\begin{figure}
    \centering
    \includegraphics[width=14cm]{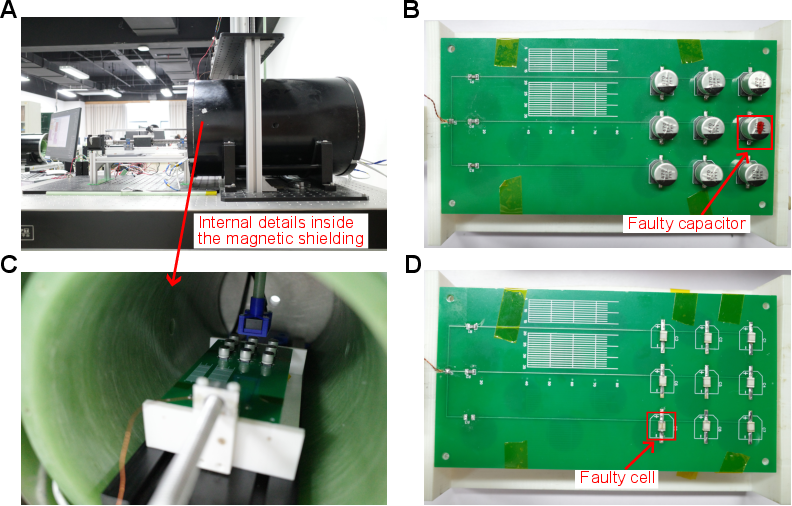}
    \caption{Experimental setup. (A) Overview of the experimental setup. The entire experimental setup is placed on an optical table. The stepper motor is placed outside the shield, and the OPM and the array of energy storage components to be detected are placed inside the shield. (C) Detailed view of the inside of the magnetic shielding. An OPM is fixed above the PCB to be tested. An array of energy storage components is connected in parallel on the PCB, which can be moved along a guiding rail. (B) Close-up view of the capacitor array under test. Nine capacitors are connected in parallel on the PCB, and the red box indicates the faulty capacitor. (D) Close-up view of the battery array under test. Nine all-solid-state batteries are connected in parallel on the PCB, and the red box indicates the faulty battery.}
    \label{fig:4}
\end{figure}

The experimental setup is shown in Figure 4. The entire experimental apparatus is placed on an optical platform (as shown in Figure~\ref{fig:4}(A)). The measurement process is conducted in a magnetically shielded environment to shield external magnetic interference. The PCB under defection and the OPM (Gen-2.0 QZFM, QuSpin) are each connected to a stepper motor (as shown in Figure~\ref{fig:4}(C)), with the former constantly moving during the measurement process to change the relative position between the PCB and OPM, and the latter used to control the distance between the OPM and the source plane prior to measurement. The capacitor array under defection is shown in Figure~\ref{fig:4}(B), with electrolytic capacitors (Panasonic 470 µF model EEEFT1E471AP). The faulty capacitor is indicated by the red box. The use of electrolytic capacitors is justified by their common usage. Therefore, studying the use of electrolytic capacitor arrays for detecting fault in capacitor arrays has practical significance, providing useful references for engineering applications. The battery array under defection is shown in Figure~\ref{fig:4}(D), with solid-state batteries (TDK model B73180A0101M199). The faulty battery is also indicated by the red box. Solid-state batteries with high energy density and high safety characteristics are expected to be widely used in portable electronic devices, medical equipment, toys, electric vehicles and other fields~\cite{li2021advance}.

\begin{figure}
    \centering
    \includegraphics[width=14cm]{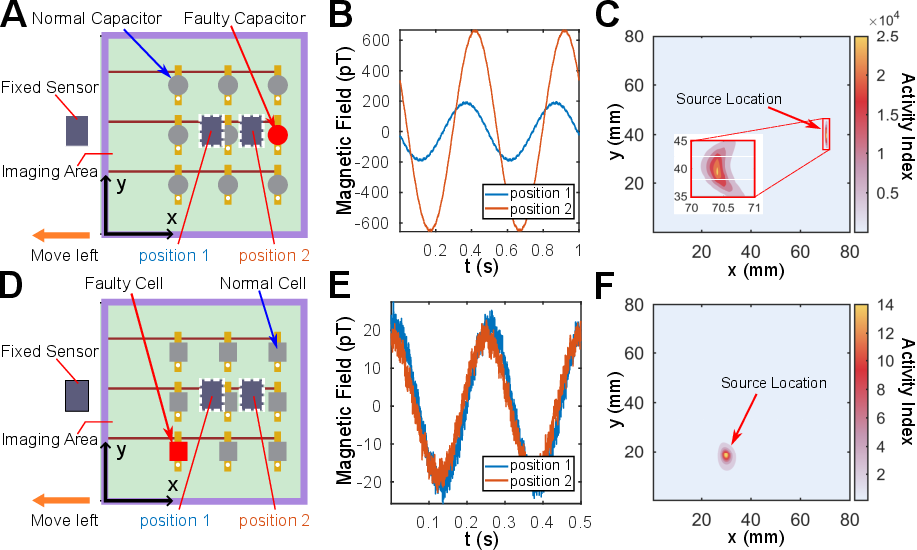}
    \caption{Top view of OPM and PCB, raw data, and imaging maps for the localization of the faulty component of the capacitor array (First row) and the battery array (Second row), respectively. (First column) Top view of the experimental setup. The light grey circles and blocks represent normal components, while the red ones represent the faulty components. Only one sensor is used in the experiment. The sensor is fixed, while the PCB is constantly moving to the left. Pos 1 and Pos 2 are the two measurement positions of the sensor. (Second column) The raw magnetic field data collected by the sensor at the two relative positions, Pos 1 and Pos 2. (Third column) Imaging maps for the localization of the faulty component.}
    \label{fig:5}
\end{figure}

In the fault localization experiment of capacitor array, an external signal source continuously charges and discharges the capacitor array at a frequency of 2 Hz. The OPM is fixed on the measurement plane 22 mm away from the surface of the PCB. The PCB travels along the x-axis with a spacing of 5 mm to traverse nine measurement points, and the magnetic field signal is collected for 1 second at each measurement point. The entire imaging area is shown as the purple rectangle in Figure~\ref{fig:5}(A), and the coordinates of the faulty capacitor in the imaging area are (70.0 mm, 40.0 mm). It should be noted that only the measurement data of the OPM's x-axis was collected in the experiment, as this is the measurement direction with the highest sensitivity of the OPM.

In the fault localization experiment of battery array, an external signal source continuously charges and discharges the battery array at a frequency of 4 Hz. The OPM is fixed on the measurement plane 7 mm away from the surface of the PCB. The PCB travels along the x-axis with a spacing of 10 mm to traverse ten measurement points, and the magnetic field signal is collected for 0.5 seconds at each measurement point. The entire imaging area is shown as the purple rectangle in Figure~\ref{fig:5}(D), and the coordinates of the faulty battery in the imaging area are (30.0 mm, 40.0 mm).

\subsection{Results of Capacitor Array Fault Localization}

Figure~\ref{fig:5}(B) shows the raw data measured by the OPM at two positions out of the nine measurement positions in Figure~\ref{fig:5}(A). It can be observed that there is a phase difference between the raw data at these two positions. This phase difference is caused by the superposition of two magnetic fields, one from the normal capacitor and the other from the short-circuited faulty capacitor. As these two magnetic fields naturally have a phase difference, the initial phases of the superimposed magnetic field at different relative positions are also different. The results in Figure~\ref{fig:5}(B) demonstrate that the OPM has captured the mixed magnetic field of the capacitor array.

Figure~\ref{fig:5}(C) shows the imaging result of the faulty capacitor, which reveals only one source on the imaging plane. The center of the source is located at the coordinates of (70.4 mm, 39.4 mm), with a localization accuracy of approximately 0.72 mm. Comparing to the less than 0.1 mm accuracy in simulation, the accuracy in experiments is lower due to several reasons, such as lower signal-to-noise ratio in experimental data, OPM not being placed completely parallel to the coordinate axis, etc.

It should be noted that no prior knowledge of the locations of the normal and faulty capacitors was known before processing the data, and only the raw data measured by the sensor as shown in Figure~\ref{fig:5}(B) was available. This further demonstrates the advantage of the proposed method, which does not require any prior knowledge of the locations of the faulty component in the array, but only relies on the sensor data and the relative positions of the sensor to the array.

\subsection{Results of Battery Array Fault Localization}

Figure~\ref{fig:5}(E) shows the raw data obtained by the OPM at the two measurement positions in Figure~\ref{fig:5}(D). Similar to Figure~\ref{fig:5}(B), there is a slight phase difference between the two sets of raw data, indicating that the OPM has detected the superimposed magnetic field from both the normal and the faulty batteries. Unlike the capacitor array, the magnetic field strength of the battery array is only about 20 pT, which is due to the use of surface-mounted solid-state batteries with lower charging and discharging currents in the experiment. In this case, the signal-to-noise ratio of the raw data is relatively low, which may affect the accuracy of fault localization. Figure~\ref{fig:5}(F) shows the imaging result of the fault localization, with the imaging position of the faulty unit marked at the coordinates of (29.9 mm, 18.4 mm), and the localization error is approximately 1.60 mm. Compared with the 0.7 mm localization error of the faulty capacitor, the localization accuracy here is lower due to the relatively low signal-to-noise ratio in the raw data.

\section{Discussion}
Currently, the time-consuming part of the detection scheme mainly lies in the need of about 10 measurement at different positions on the measurement plane, with each of them lasting for 1 second. In the faulty battery cell detection experiments, an attempt was made to increase the charging and discharging frequency from 2Hz to 4Hz, which reduced the measurement time at each position from 1 second to 0.5 seconds, as shown in Figure~\ref{fig:5}(E). The imaging results were almost unaffected. Therefore, increasing the charging and discharging frequency can increase the detection speed.

On the other hand, this paper only discusses the localization results of having a single short-circuit faulty component inside an array. Subsequent work will focus on the localization of multiple faulty components in an energy storage component array, possibly with different faults or different degrees of faults.

This method can be further extended to fault localization of other electronic component arrays. By following the procedure described in Equation~(\ref{eq1}) to establish the relationship between the intensity of the target fault source under detection and the magnetic field distribution in its surrounding space, the lead field matrix represented in Equation~(\ref{eq2}) can be derived. By applying the methodology presented in this paper, the localization of the faulty electronic component can be achieved.

\section{Conclusion}
This paper proposes a novel method for fault localization of energy storage component arrays, using a single highly sensitive magnetometer to measure the magnetic field around the arrays. The proposed method has been successfully applied to the fault detection of capacitor and solid-state battery arrays in experiments, achieving precise localization results. The designed PCB circuit board ensures that the sensor does not measure the magnetic field generated by external currents on the circuit board. Moreover, the robustness of the proposed method is demonstrated through an impact analysis of different signal-to-noise ratios in measurements to the localization accuracy. The results have indicated that the method can achieve accurate imaging even in low signal-to-noise ratio scenarios. The proposed method can also be scaled and applied to fault diagnosis of other electronic component arrays, such as inductor arrays. Furthermore, due to the utilization of a single sensor and a mobile platform in the proposed method, it can be easily integrated into industrial production processes for real-time and precise localization of faulty products during the manufacturing process.


\section*{Acknowledgements}
N.Y.H. gratefully acknowledge the support provided by the China National Key R\&D Program (project number 2022YFA1604502). WG gratefully acknowledge the support provided by National Natural Science Foundation of China under U22B2040. SZ and WG gratefully acknowledge the support provided by National Natural Science Foundation of China under U21A20119.



\bibliographystyle{elsarticle-num} 

\bibliography{ref}





\end{document}